# Characteristics of Compound Multiplicity in $^{84}Kr_{36}$ with various light and heavy targets at 1 GeV per nucleon

N S Chouhan, M K Singh and V Singh*
*Nuclear and Astroparticle Physics Laboratory, Department of Physics,*
*Banaras Hindu University, Varanasi – 221005, India.*
* Email: venkaz@yahoo.com and corresponding author

**Abstract**

Present article focuses on the interactions of $^{84}Kr_{36}$ having kinetic energy around 1 GeV per nucleon with NIKFI BR-2 nuclear emulsion detector's targets, that can reveal important features of some compound multiplicity. The observation showed that the width of the compound multiplicity distributions and value of the mean compound multiplicity have linear relation with the mass number of the projectile as well as colliding system.

**Keywords:** Projectile fragmentation, Compound Multiplicity, Nuclear Emulsion detector.

**PACS:** 29.40.Rg; 25.70.Mn, 25.70.Pq; 25.75.Dw;

## 1. Introduction

The investigation of final state particles produced through nucleon – nucleus or nucleus – nucleus interactions at high energy is an active research area with discovery potential. The nuclear emulsion detector (NED) is one of the oldest detector technologies and has been in use from the birth of the experimental nuclear and astroparticle physics. The interaction of hadrons (such as $^{84}Kr_{36}$) with nuclear emulsion detector's target at relativistic energy reveals the picture of the hadrons – nucleus collisions. The recoil target nucleons are emitted shortly after the passage of the leading hadrons. Therefore, it is worthy to study the compound multiplicity of recoil target nucleon and freshly produced particles mainly pions during the interactions as it may contain important information about the interaction. The number of target's recoil nucleon and produced charged particles have been taken together and termed as the compound multiplicity. Most of the experiments on high energy collisions have been carried out to investigate the characteristics of the newly created particles mostly pion, meson and Kaons (shower; $N_s$) and recoiled target's nucleon (grey particles; $N_g$). The study of the characteristic of grey particles produced in such collision has also increased due to the fact that their emission takes place on the time scale of same order (~$10^{-22}$ s) as that of shower particles. Hence, they are expected to keep some memory of the reaction history [1]. Furthermore, the grey particle may also be taken as a good measure of the number of encounters made by the impinging hadrons inside the struck nucleons [2, 3]. In order to refine the models for multiparticle production in hadron-nucleus and nucleus-nucleus collisions, a new variable termed as compound multiplicity ($N_c = N_g + N_s$), was introduced by Jurak and Linscheid [2 - 8] in year 1977 and some interesting characteristics of grey and shower particles, taken together per interactions, have been investigated by several workers [2 - 8]. Relativistic high energy interactions of various heavy ion (projectile) [9-12] with emulsion have been studied. Therefore, it is interesting to study the use of the compound multiplicity of grey and shower particles. The present work is devoted mainly to the discussion of the experimental data on the compound multiplicity distribution and their characteristic with respect to other emitting particles in the inelastic collision of $^{84}$Kr with the nuclei of the nuclear emulsion detector at around 1 GeV per nucleon.

## 2. Experimental details:

Nuclear emulsion detector is composed of silver halide crystals immersed in a gelatin matrix [2, 13-18] consisting mostly of hydrogen, carbon, nitrogen, oxygen, silver and bromine while a small percentage of sulfur and iodine are also present. In the present experiment, we have employed a stack of high sensitive NIKFI BR-2 nuclear emulsion pellicles of dimensions 9.8×9.8×0.06 $cm^3$, exposed horizontally to $^{84}Kr_{36}$ ion at kinetic energy of around 1 GeV per nucleon. The exposure has been performed at Gesellschaft fur Schwerionenforschung (GSI) Darmstadt, Germany. Interactions were found by along-the-track scanning technique using an oil immersion objective of 100X magnification. The beam tracks were picked up at a distance of 5 mm from the edge of the plate and carefully followed until they either interacted with nuclear emulsion detector nuclei or escaped from any surface of the emulsion. A total of 700 inelastic events produced in $^{84}$Kr- emulsion interactions were located. In the present analysis, out of 700 there are 570 events having fulfilled the required criteria of further investigation [13-18]. All charge secondary emitted or produced in an interaction are classified in accordance with their ionization, range and velocity into the following categories [13-18]:

***(a) Shower tracks ($N_s$):*** These are freshly created charged particles with normalized grain density or ionization $g^* < 1.4$ ($g^*$ is the normalized grain density). These particles have relative velocity ($\beta$) > 0.7. For the case of a proton, kinetic energy ($E_p$) should be less than 400 MeV.

They are mostly fast pions with a small admixture of Kaons and released protons from the projectile which have undergone an interaction.

***(b) Grey tracks ($N_g$):*** Particles having ionization in the interval $1.4 < g^* < 6.0$ and range $L > 3$ mm are defined as greys. This particle has relative velocity ($\beta$) in between $0.3 < (\beta) < 0.7$. They are generally knocked out protons (nuclear emulsion detector is not able to detector neutral particle) of targets having kinetic energy in between $30 < E_p < 400$ MeV but also admixture of deuterons, tritons and some slow mesons.

***(c) Black tracks ($N_b$):*** Particles having range $L < 3$ mm from interaction vertex from which they originated and $g^* > 6.0$. This corresponds to a relative velocity $\beta < 0.3$ and a proton with kinetic energy $E_p < 30$ MeV. Most of these are produced due to evaporation of residual target nucleus.

Nuclear emulsion detector is a composite target detector and major target components are H, C, N, O, and Ag and Br. The number of heavily ionizing charged particles ($N_h = N_b + N_g$) depends upon the target breakup. Therefore we used such distribution to fix criteria for target separation [13-18].

***Ag / Br target events:*** $N_h \geq 8$ and at least one track with $R < 10$ μm is present in an event. This class of target was the cleanest interaction and high statistics are known as Ag/Br target events.

***CNO target events:*** $1 < N_h < 8$ and no tracks with $R < 10$ μm are present in an event. This class always contains very clean interaction of CNO target.

***H target events:*** $N_h \leq 1$ and no tracks with $R < 10$ μm are present in an event. This class include all $^{84}$Kr + H interactions but also some of the peripheral interactions with CNO and the very peripheral interactions with Ag / Br targets.

On the basis of the above criteria, we obtained 13.4, 39.0 and 47.6 percent of interactions with H, CNO and Ag / Br targets, respectively [13-18].

## 3. Results and Discussion:

Figure 1 represents the compound multiplicity distributions for $^{84}Kr_{36}$ interactions with different target groups of nuclear emulsion detector such as H, CNO and Ag/Br, selected on the basis of heavily ionizing charged particle ($N_h$). Figure 1 reveals that the multiplicity distributions become wider with increasing target size. The average value of compound multiplicity $\langle N_c \rangle$, its dispersion $D(N_c) = \sqrt{(\langle N_c^2 \rangle - \langle N_c \rangle^2)}$ and the ratio $\langle N_c \rangle / D(N_c)$ are presented in Table 1. Figure 2 shows the dependence of the average compound multiplicity $\langle N_c \rangle$ on the mass number of the beam nucleus $A_p$. It may be observed from the figure 2 that $\langle N_c \rangle$ increase rapidly with increasing mass of the incident beam. The points are the experimental data while the continuous line is the result of fitting by the relation $\langle N_c \rangle = K^{\alpha}{}_{\beta}$.

The average multiplicities of black, grey and shower particles in different ensembles of $^{84}Kr$ – nuclear emulsion interactions are presented in table 2. For comparison, corresponding results from $^{84}Kr$ – nuclear emulsion collisions and $^{56}Fe$ – nuclear emulsion collisions at 1.7 GeV per nucleon are also tabulated in table 2. From table 2, we can clearly see that the average multiplicity of black, grey and shower particles increases in the average target mass for both $^{84}Kr$ – nuclear emulsion and $^{56}Fe$ – nuclear emulsion interactions.

Koba - Nielsen - Olesen (KNO) scaling is a well established empirical law for multiparticle production in p + p collision [13-18]. The KNO scaling behavior of compound multiplicity distributions is observed in the above emulsion data. Fig. 3 shows the $\langle N_c \rangle( \sigma_n / \sigma_{inel})$ versus $N_c/\langle N_c \rangle$ for $^{84}Kr$-emulsion interactions at 1 GeV per nucleon, where $\sigma_n$ denotes the partial cross section for producing n charged compound particles, $\sigma_{inel}$ denotes the total inelastic cross section, and $z = N_c/\langle N_c \rangle$, respectively. From figure 3, it can be seen that the experimental data lie on a universal curve, which can be fitted by a KNO scaling function with the form

$$\Phi(z) = (Az + Bz^3 + Cz^5 + Dz^7)e^{Ez}. \quad (1)$$

The best fitting parameters are A = 8.34 ± 1.28, B = - (1.46 ± 2.38), C = 3.99 ± 2.76, D = 1.17 ± 0.77 and E = - (3.41 ± 0.88). The minimum fitting $\chi^2$ / DOF is 1.05, which means that the experimental data can be well explained by the KNO scaling law. The fitting parameter value is similar as reported in [2] for $^{84}Kr$ – emulsion interaction at 1.7 GeV per nucleon.

Figure 4 shows the correlations between $\langle N_c \rangle$ and $N_s$, $N_h$, $N_g$, $N_b$ for $^{84}$Kr –emulsion collisions at around 1 GeV per nucleon. These correlations are nicely fitted by a linear relation of the form,

$$\langle N_c \rangle = (1.01 \pm 0.04) N_b + (3.28 \pm 0.67), \quad (2)$$

$$\langle N_c \rangle = (0.95 \pm 0.03) N_g + (6.70 \pm 0.66), \quad (3)$$

$$\langle N_c \rangle = (0.63 \pm 0.03) N_s + (5.07 \pm 0.77), \quad (4)$$

$$\langle N_c \rangle = (0.96 \pm 0.03) N_h + (4.87 \pm 0.69). \quad (5)$$

Figure 5 shows the correlations between $\langle N_s \rangle$, $\langle N_h \rangle$, $\langle N_g \rangle$, $\langle N_b \rangle$ and $\langle N_c \rangle$ for $^{84}$Kr – nuclear emulsion collisions at around 1 GeV per nucleon. These correlations are nicely fitted by a linear relation of the form,

$$\langle N_b \rangle = (0.28 \pm 0.05) N_c + (0.72 \pm 0.07), \quad (6)$$

$$\langle N_g \rangle = (0.41 \pm 0.03) N_c + (1.85 \pm 0.76), \quad (7)$$

$$\langle N_s \rangle = (0.59 \pm 0.03) N_c + (4.21 \pm 0.77), \quad (8)$$

$$\langle N_h \rangle = (0.43 \pm 0.02) N_c + (2.58 \pm 0.69). \quad (9)$$

## 4. Conclusions:

The present study reveals that the compound particle multiplicity distribution increases rapidly with increase in mass number of the incident projectile. The multiplicity correlation between $N_b$, $N_g$, $N_s$, $N_h$, and average value of $N_c$ is showing linear relation. The compound multiplicity distribution is observed to obey a KNO scaling law, which is independent of the mass number and energy of the projectile.

**Acknowledgement:** Authors are grateful to the all technical staff of GSI, Germany for exposing nuclear emulsion detector with $^{84}Kr_{36}$ beam.

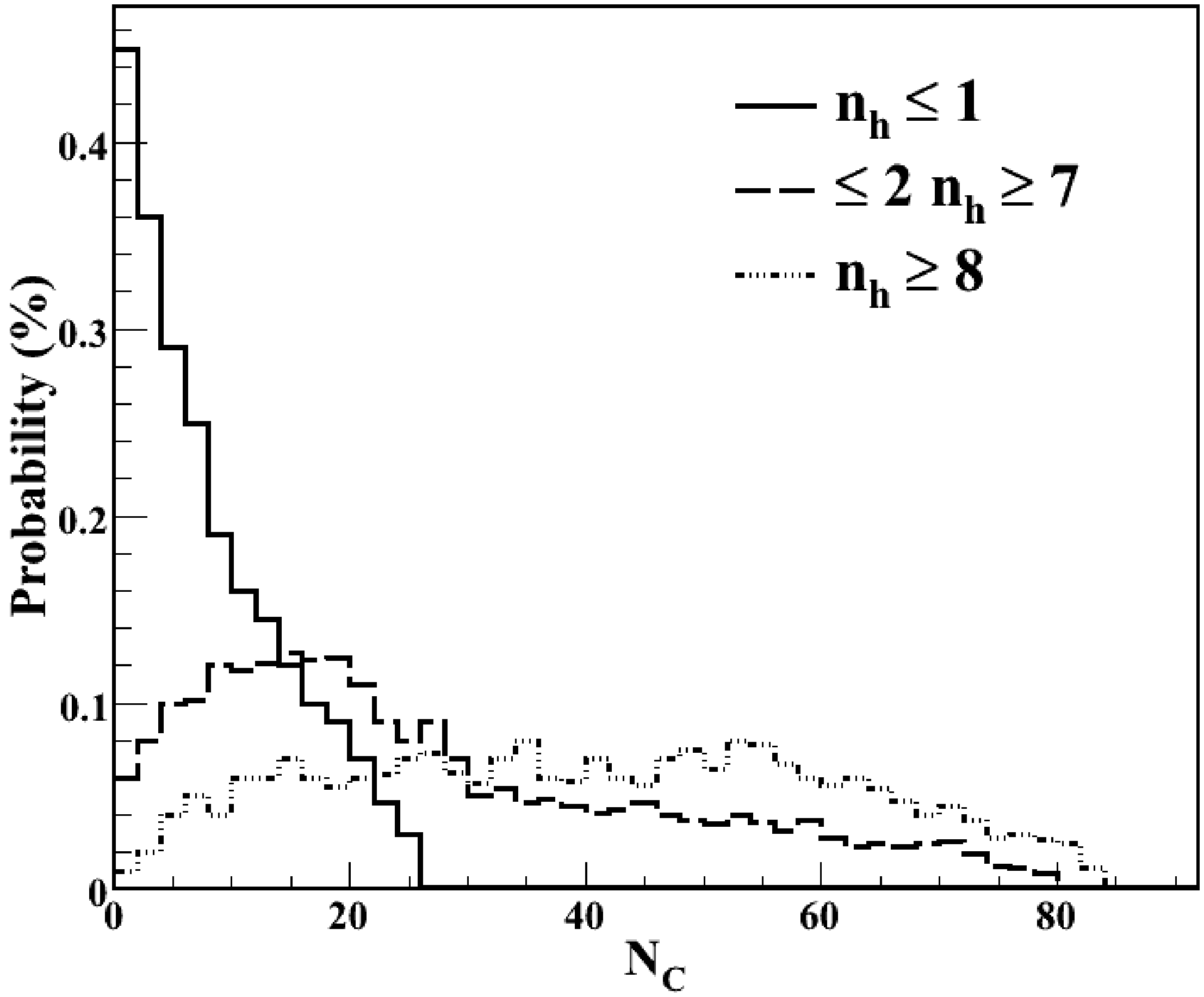

**Fig. 1:** Compound multiplicity distributions for different emulsion target groups in $^{84}Kr_{36}$ with nuclear emulsion collisions at ~1 GeV per nucleon.

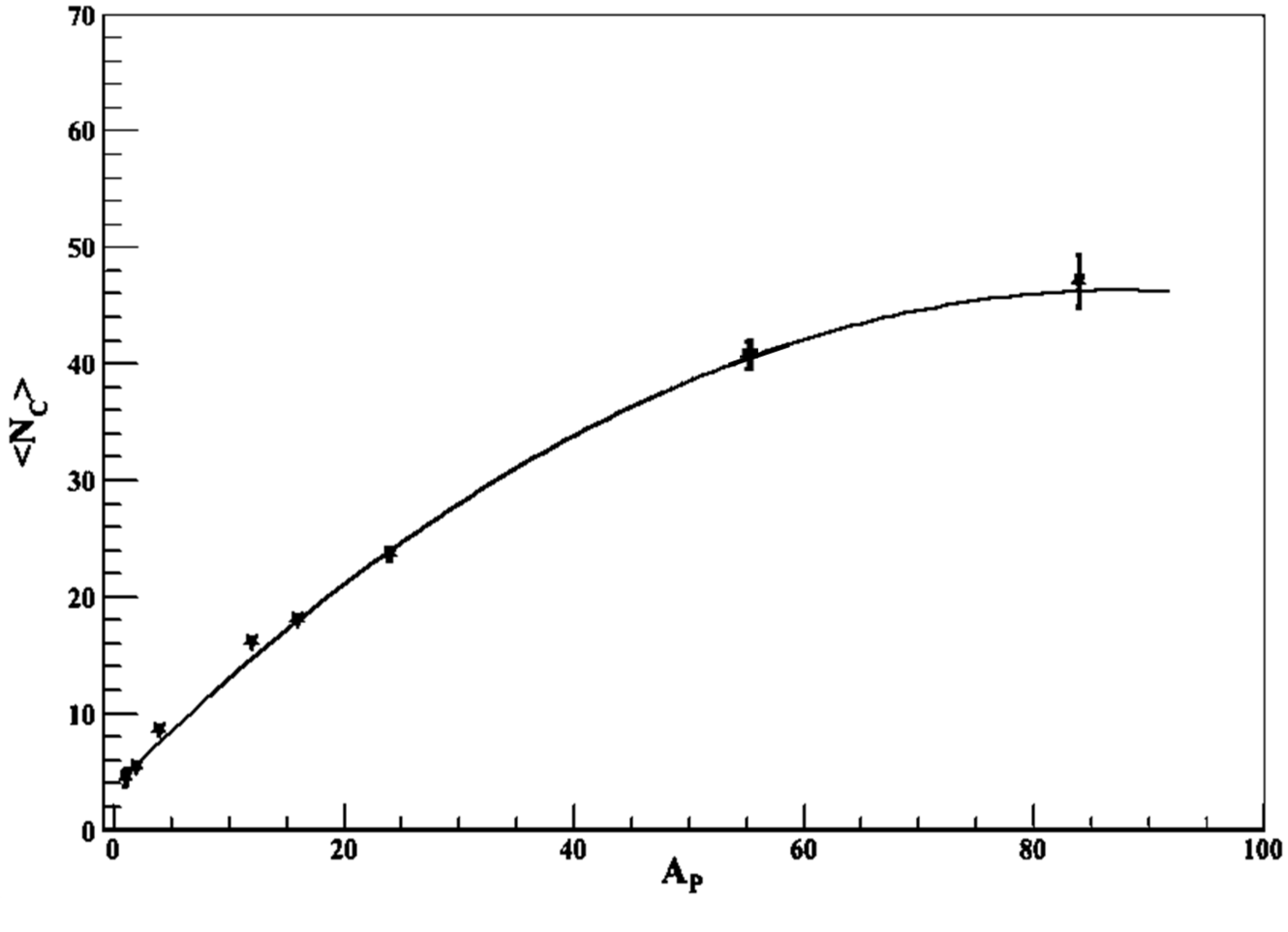

**Fig. 2:** Dependence of <$N_c$> on the mass number $A_p$ of different projectile in nucleus – nucleus collisions.

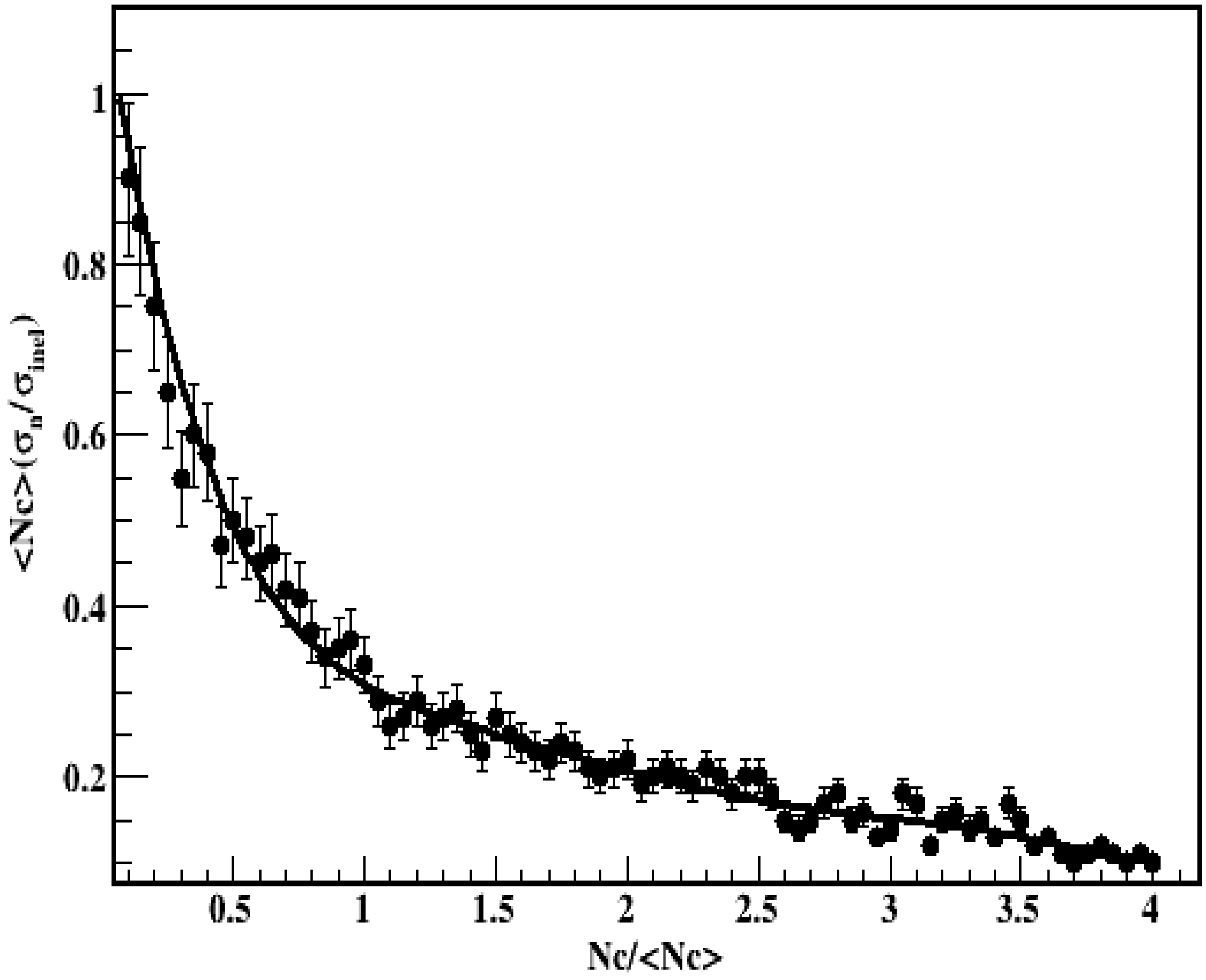

**Fig. 3:** $<N_c>(\sigma_n / \sigma_{inel})$ versus $N_c/<N_c>$ for $^{84}$Kr – nuclear emulsion interactions at 1 GeV per nucleon.

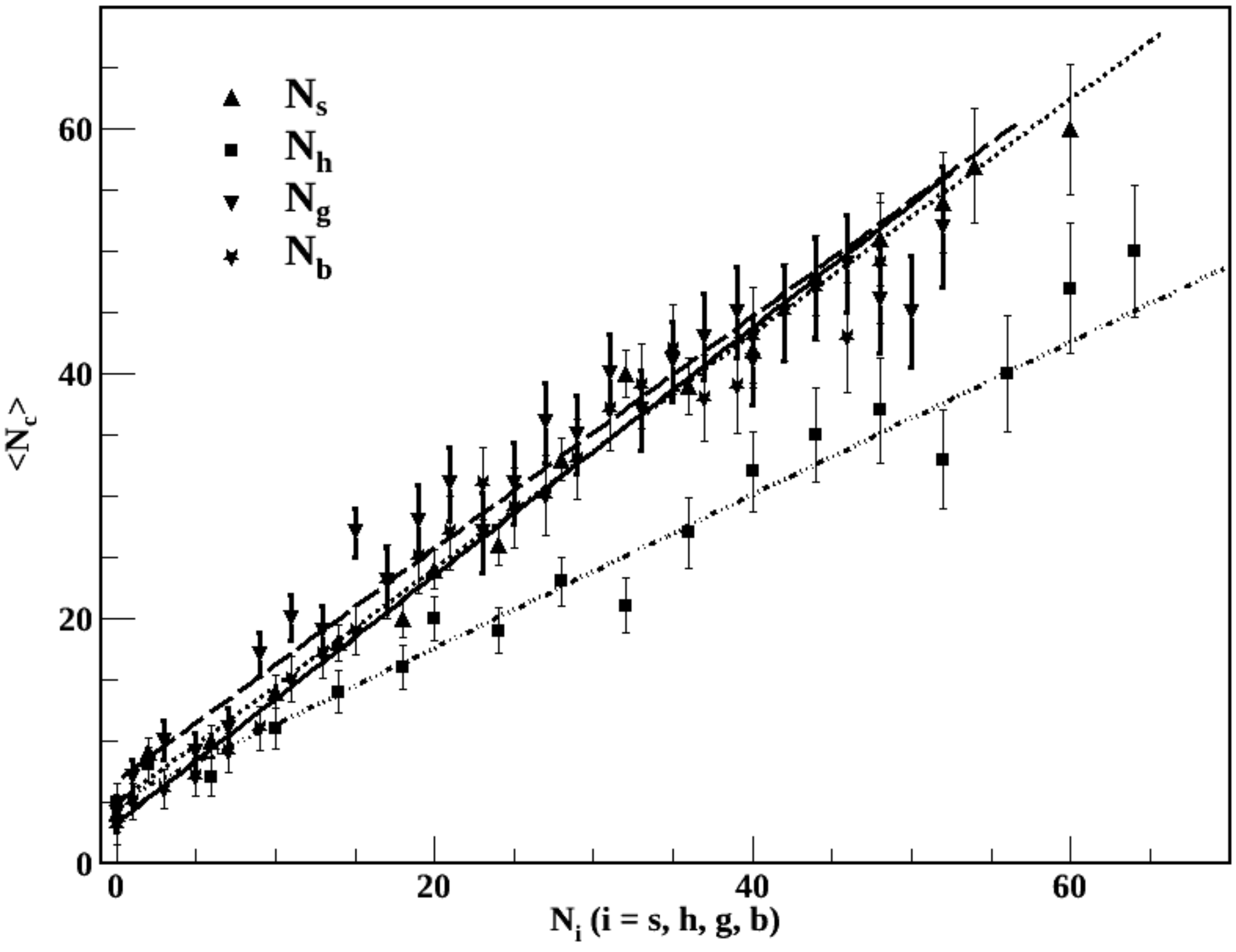

Fig. 4: Dependence of $<N_c>$ on $N_i$ (i = s, h, g, b) for $^{84}$Kr with emulsion at ~1 GeV per nucleon.

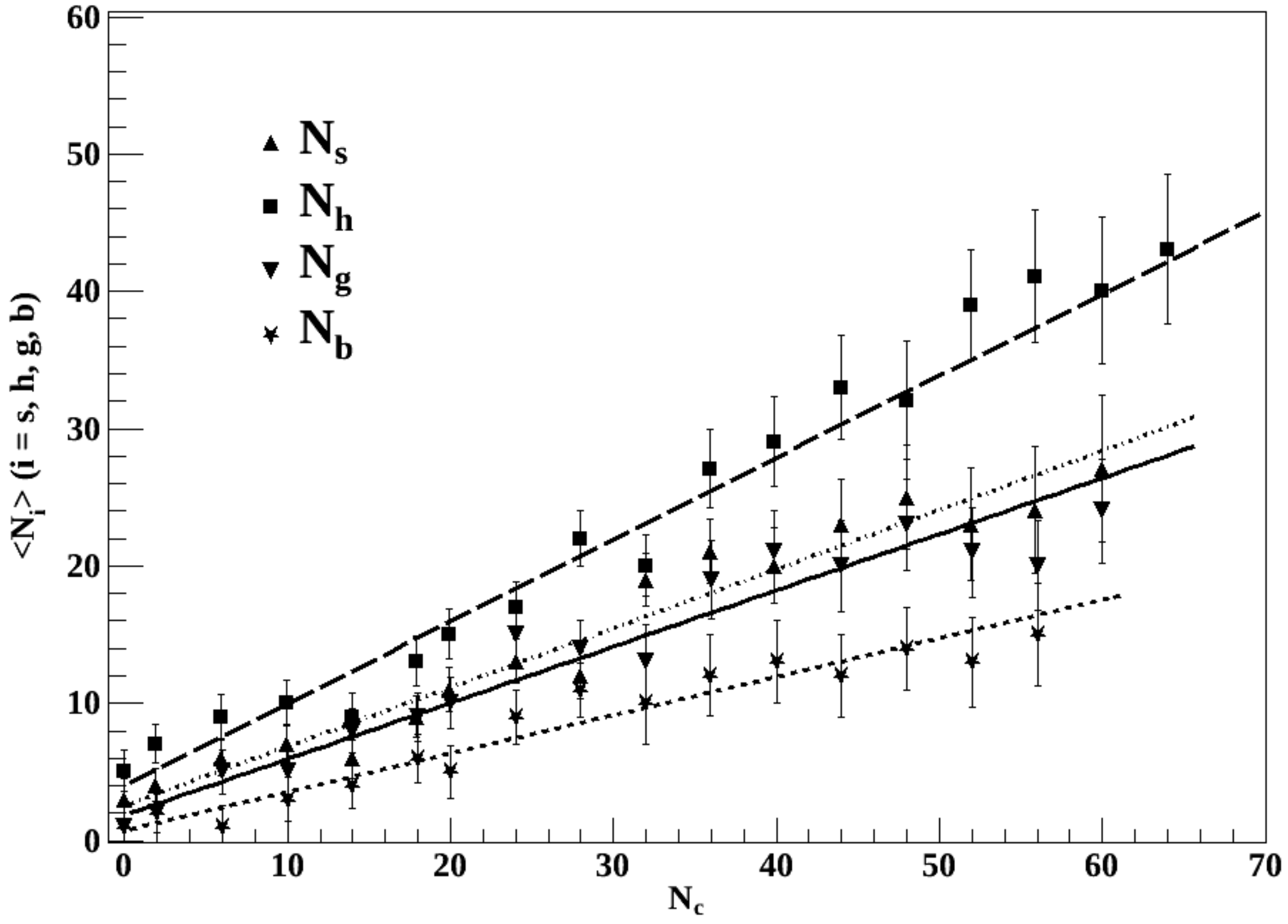

**Fig. 5:** Dependence of $\langle N_b \rangle$ and $\langle N_h \rangle$ on $N_c$ for $^{84}$Kr with emulsion at ~1 GeV per nucleon.

**Table 1:** The value of the mean compound multiplicity $\langle N_c \rangle$, dispersion $D(N_c)$ and ratio $\langle N_c \rangle$ / $D(N_c)$ in $^{84}$Kr interactions.

| **Target group** | **$\langle N_c \rangle$** | **$D(N_c)$** | **<Nc>/D(Nc)** |
|---|---|---|---|
| **H** | 7.56 ± 0.12 | 4.47 ± 0.32 | 1.69 ± 0.37 |
| **CNO** | 14.93 ± 0.15 | 10.66 ± 0.28 | 1.40 ± 0.53 |
| **Emulsion** | 47.04 ± 0.27 | 27.19 ± 0.40 | 1.73 ± 0.67 |
| **Ag/Br** | 53.27 ± 0.31 | 20.10 ± 0.79 | 2.65 ± 0.39 |

**Table2:** The percentage occurrence and average multiplicities of charged particles in different target groups of $^{84}Kr$ (1.7 A GeV), $^{56}Fe$ (1.7 A GeV) and $^{84}Kr$ (~ 1 A GeV) interactions.

| **Beam** | **Type of events** | **Percentage** | **<$N_b$>** | **<$N_g$>** | **<$N_s$>** | **Ref.** |
|---|---|---|---|---|---|---|
| $^{56}Fe$ | $N_h \leq 1$ | 14.33±1.24 | 0.15±0.03 | 0.32±0.05 | 2.77±0.20 | [19] |
| $^{56}Fe$ | $1 < N_h < 8$ | 34.12±1.91 | 1.84±0.07 | 2.91±0.13 | 8.03±0.43 | [19] |
| $^{56}Fe$ | $N_h \geq 8$ | 51.55±2.35 | 7.38±0.18 | 14.90±0.50 | 19.60±0.60 | [19] |
| $^{56}Fe$ | $N_h \geq 0$ | 100 | 4.45±0.14 | 8.71±0.34 | 13.30±0.40 | [19] |
| $^{84}Kr$ | $N_h \leq 1$ | 16.67±1.73 | 0.08±0.03 | 0.19±0.04 | 4.68±0.79 | [2] |
| $^{84}Kr$ | $1 < N_h < 8$ | 41.76±2.74 | 2.17±0.11 | 1.97±0.10 | 7.08±0.48 | [2] |
| $^{84}Kr$ | $N_h \geq 8$ | 41.58±2.73 | 10.22±0.29 | 9.91±0.44 | 14.69±0.77 | [2] |
| $^{84}Kr$ | $N_h \geq 0$ | 100 | 5.17±0.22 | 4.98±0.26 | 9.84±0.44 | [2] |
| $^{84}Kr$ | Nh≤1 | 14.76±1.23 | 0.06±0.03 | 0.19±0.06 | 3.99±0.69 | **Present work** |
| $^{84}Kr$ | 1 < Nh < 8 | 41.85±2.78 | 1.99±0.29 | 1.75±0.47 | 8.10±0.90 | **Present work** |
| $^{84}Kr$ | Nh ≥ 8 | 43.07±1.97 | 10.05±0.21 | 10.09±0.38 | 15.09±0.60 | **Present work** |
| $^{84}Kr$ | Nh ≥ 0 | 100 | 6.25±0.22 | 5.21±0.47 | 8.99±0.73 | **Present work** |